\documentclass[a4]{article}
\usepackage{amssymb}
\usepackage{amsmath}
\usepackage{latexsym}
\usepackage{array}
\usepackage{bm}
\usepackage{theorem}
\usepackage{graphicx}
\usepackage{exscale}

\theoremstyle{break}
\newtheorem{Theorem}{Theorem}[section]

\newtheorem{Proposition}[Theorem]{Proposition}
\newtheorem{Lemma}[Theorem]{Lemma}
\newtheorem{Remark}[Theorem]{Remark}
\newtheorem{Definition}[Theorem]{Definition}
\def\DIS{\displaystyle}
\def\hbreak{\vspace*{2mm}\hfill\break\noindent}

\def\n{\boldsymbol{n}}
\def\r{\boldsymbol{r}}
\def\x{\boldsymbol{x}}

\def\p{\boldsymbol{p}}
\def\q{\boldsymbol{q}}
\def\u{\boldsymbol{u}}
\def\e{\boldsymbol{e}}
\def\b0{\boldsymbol{0}}
\def\btau{\boldsymbol{\tau}}
\def\bT{\boldsymbol{T}}
\def\bF{\boldsymbol{F}}
\def\bU{\boldsymbol{U}}
\def\bW{\boldsymbol{W}}

\def\Z{\mathbb{Z}}
\def\Q{\mathbb{Q}}

\def\mR{\mathcal{R}}
\def\mK{\mathcal{K}}
\def\mS{\mathcal{S}}
\def\mI{\mathcal{I}}

\def\ex{\mbox{e}}

\begin{document}

\title{Nonlinear forms of coprimeness preserving extensions to the Somos-$4$ recurrence and the two-dimensional Toda lattice equation --investigation into their extended Laurent properties--}
\author{Ryo Kamiya$^1$, Masataka Kanki$^2$, Takafumi Mase$^1$, Tetsuji Tokihiro$^1$\\
\small $^1$ Graduate School of Mathematical Sciences,\\
\small the University of Tokyo, 3-8-1 Komaba, Meguro, Tokyo 153-8914, Japan\\
\small $^2$ Department of Mathematics, Faculty of Engineering Science,\\
\small Kansai University, 3-3-35 Yamate, Suita, Osaka 564-8680, Japan}

\date{}

\maketitle

\begin{abstract}
Coprimeness property was introduced to study the singularity structure of discrete dynamical systems.
In this paper we shall extend the coprimeness property and the Laurent property to further 
investigate discrete equations with complicated pattern of singularities.
As examples we study extensions to the Somos-$4$ recurrence and the two-dimensional discrete Toda equation.
By considering their non-autonomous polynomial forms, we prove that their tau function analogues possess the extended Laurent property with respect to their initial variables and some extra factors related to the non-autonomous terms.  
Using this Laurent property, we prove that these equations satisfy the extended coprimeness property.
This coprimeness property reflects the singularities that trivially arise from the equations.
\end{abstract}

\section{Introduction}
The Toda lattice is a system of differential equations invented by Morikazu Toda in 1967 as a model of a linear chain of masses interacting with each other according to an exponential interaction force \cite{Toda}.
A two-dimensional extension was achieved by A.\ V.\ Mikhailov, and later several forms of two-dimensional Toda lattices were proposed \cite{Mikhairov,Fordy-Gibbons}.
In this article we shall pick up a hyperbolic partial differential equation form
\begin{equation}\label{2Dtoda_eq}
\frac{\partial^2}{\partial x\partial y}\log U_t(x,y)=U_{t+1}(x,y)-2U_t(x,y)+U_{t-1}(x,y),
\end{equation}
and study the properties of discrete analogues (and their extensions) of the equation \eqref{2Dtoda_eq} in relation to the integrability of the equations as discrete dynamical systems.

From the 1970s, discrete analogues of integrable systems began to draw keen attention, partly due to Ryogo Hirota's works on integrable discretizations of various integrable systems, such as the (one-dimensional) discrete Toda lattice
\cite{Hirota}. Later in \cite{HTI,Tsujimoto} a discrete analogue of the two-dimensional Toda lattice was introduced as the following recurrence:
\begin{equation}\label{nonlinear_Toda1}
\frac{(U_{t+1,n,m+1}-1)(U_{t-1,n+1,m}-1)}{(U_{t,n+1,m}-1)(U_{t,n,m+1}-1)}
=\frac{U_{t,n,m}U_{t,n+1,m+1}}{U_{t,n+1,m}U_{t,n,m+1}}.
\end{equation}
Let us briefly review how \eqref{nonlinear_Toda1} corresponds to the equation \eqref{2Dtoda_eq} through a continuous limit.
If we define
\[
\tilde{U}_{t,n,m}:=\frac{-1+\delta \epsilon}{\delta \epsilon}U_{t,n,m},
\]
the equation \eqref{nonlinear_Toda1} turns into
\[
\frac{\left(1+\delta \epsilon(\tilde{U}_{t+1,n,m+1}-1)\right)\left(1+\delta \epsilon(\tilde{U}_{t-1,n+1,m}-1)\right)}{\left(1+\delta \epsilon(\tilde{U}_{t,n+1,m}-1)\right)\left(1+\delta \epsilon(\tilde{U}_{t,n,m+1}-1)\right)}
=\frac{\tilde{U}_{t,n,m}\tilde{U}_{t,n+1,m+1}}{\tilde{U}_{t,n+1,m}\tilde{U}_{t,n,m+1}},
\]
where we denoted $\tilde{U}_{t,n,m}:=U_t(n \delta, m \epsilon)$.  
By fixing the values $x:=n \delta$ and $y:=m \epsilon$ and taking the limits $\delta \rightarrow 0$ and $\epsilon \rightarrow 0$, we
recover  \eqref{2Dtoda_eq}.

Another form of discretizations of \eqref{2Dtoda_eq} is
\begin{equation}
\tau_{t+1,n,m+1}\tau_{t-1,n+1,m}=\tau_{t,n+1,m}\tau_{t,n,m+1}+\tau_{t,n,m}\tau_{t,n+1,m+1},
\label{2DdiscreteToda_eq}
\end{equation}
which is called the bilinear form of the two-dimensional discrete Toda equation, and $\tau_{t,n,m}$ is called  the ``tau-function''
of the two-dimensional discrete Toda equation.
The equation \eqref{2DdiscreteToda_eq} was also proposed by Hirota and his collaborators \cite{HTI, Tsujimoto}.
The continuous equation \eqref{2Dtoda_eq} looks quite different in shape from the discrete one \eqref{nonlinear_Toda1}, however, it is known that by using the bilinear form as an intermediary, the correspondence between the continuous and the discrete is naturally understood.

As various discrete analogues of integrable systems began to arise, elaboration on the ``definition'' of integrability of discrete equations has been needed.
One of the most famous criteria of discrete integrability is the singularity confinement test \cite{SC}, which was invented as an analogy to the Painlev\'{e} test for ordinary differential equations.
Another one is the algebraic entropy test \cite{AE}, which asserts that an equation is considered to be integrable if the degree growth of its iterates is below exponential order (in other words polynomial order).
Both of them have been so useful in distinguishing integrable systems from nonintegrable ones that most of the known equations have been studied in terms of these tests.
In the course of these investigations some difficulties are found out.
One of them is the ``contradiction'' between these two criteria such as in \cite{HV}.
Another one is the difficulty of application of the singularity confinement to discrete equations over a higher dimensional lattice.
When we investigate the singularity patterns of a discrete equation over multi-dimensional lattices, we encounter a problem:
the number of patterns of the singularities arising from the configurations of the initial variables is so large that we can no longer 
pick up all the singularities and test whether they are confined.
To resolve this kind of difficulty, we introduced the coprimeness property of discrete equations.
\begin{Definition} \label{def11}
Let $\{x_{\n}\}$ be a discrete dynamical system with the independent variable $\n$ and the dependent variable $x$.
\begin{itemize}
\item $\{x_{\n}\}$ has the Laurent property if each iterate of the equation is expressed as a Laurent polynomial of the initial variables.
\item $\{x_{\n}\}$ has the coprimeness property if there exist a positive constant $D$ such that for every pair of iterates $x_{\n}, x_{\n'}$ with $d(\n,\n')\ge D$, they have a decomposition $x_{\n}=f_1/f_2$, $x_{\n'}=g_1/g_2$ where $f_1,f_2,g_1,g_2$ are polynomials of the initial variables and any common factor of arbitrary two elements in $\{f_1,f_2,g_1,g_2\}$ is a monomial.
\end{itemize}
\end{Definition}
Here we assume that we have a suitable metric $d(\cdot,\cdot)$ on the lattice (but the definition of the coprimeness property does not depend on the choice of $d(\cdot,\cdot)$).
Our claim from several examples is that the coprimeness property is an algebraic reinterpretation  (and also a slight generalization) of singularity confinement.
Let us summarize our previous studies on the coprimeness property.
The coprimeness of the discrete KdV equation was formulated and proved in \cite{KMMT2}.
In \cite{dToda}, Mada and two of the authors investigated initial value dependence of the discrete Toda lattice equation with several boundary conditions, and showed that the discrete Toda lattice has the coprimeness property.
In the subsequent paper \cite{KKMT}, we introduced an extension of the bilinear two-dimensional discrete Toda equation:
\begin{equation}
\tau_{t+1,n,m+1}\tau_{t-1,n+1,m}=\tau_{t,n+1,m}^{k_1}\tau_{t,n,m+1}^{k_2}+\tau_{t,n,m}^{l_1}\tau_{t,n+1,m+1}^{l_2},
\label{pDToda_polinear_eq}
\end{equation}
where $k_1,k_2,l_1,l_2$ are positive integers.
For $k_1=k_2=l_1=l_2=1$, \eqref{pDToda_polinear_eq} is equivalent to the classical \eqref{2DdiscreteToda_eq}.
We proved that $\tau_{t,n,m}$ are mutually independent irreducible Laurent polynomials with respect to the initial variables $\{ \tau_{0,n,m},\,\tau_{1,n,m}\}_{n,m\in\Z}$ as far as the greatest common divisor GCD$(k_1,k_2,l_1,
l_2)$ of the indices is a non-negative power of two, which means that the right hand side of \eqref{pDToda_polinear_eq} is irreducible.

In this article, we report on the following discrete lattice equation
\begin{equation}\label{nonlinear_pseudo_Toda1}
\frac{(U_{t+1,n,m+1}-1)(U_{t-1,n+1,m}-1)}{(U_{t,n+1,m}-1)^{k_1}(U_{t,n,m+1}-1)^{k_2}}
=\frac{U_{t,n,m}^{l_1}U_{t,n+1,m+1}^{l_2}}{U_{t,n+1,m}^{k_1}U_{t,n,m+1}^{k_2}},
\end{equation}
where $t \in \Z_{\ge 0}$, $(n,m) \in \Z^2$ and the parameters $k_1,k_2,l_1,l_2$ are arbitrary positive integers.
Here we think of $t$ as a time variable, and consider the time evolution of \eqref{nonlinear_pseudo_Toda1} for $t\ge 0$.
When $k_1=k_2=l_1=l_2=1$, \eqref{nonlinear_pseudo_Toda1} coincides with \eqref{nonlinear_Toda1}, the nonlinear form of the  two-dimensional Toda lattice equation.
The equation \eqref{nonlinear_pseudo_Toda1} is obtained as an extension to \eqref{nonlinear_Toda1} corresponding to the extended ``bilinear'' form \eqref{pDToda_polinear_eq}.
The transformation from \eqref{pDToda_polinear_eq} to \eqref{nonlinear_pseudo_Toda1} is fairly simple.
For a given solution of \eqref{pDToda_polinear_eq}, if we put 
\begin{equation}\label{UtoTau}
U_{t,n,m}:= \frac{\tau_{t+1,n,m+1}\tau_{t-1,n+1,m}}{\tau_{t,n+1,m}^{k_1}\tau_{t,n,m+1}^{k_2}}, 
\end{equation}
$U_{t,n,m}$  satisfies \eqref{nonlinear_pseudo_Toda1}.
Hence one might expect that \eqref{nonlinear_pseudo_Toda1} also has the coprimeness property and that this conjecture will be proved easily from the irreducibility of \eqref{pDToda_polinear_eq}. 
However, for a solution of \eqref{nonlinear_pseudo_Toda1}, $\tau_{t,n,m}$ in \eqref{UtoTau} does not necessarily satisfy \eqref{pDToda_polinear_eq}.
The equation for $\tau_{t,n,m}$ depends on the initial values of $U_{t,n,m}$ and we cannot directly use the irreducibility of $\tau_{t,n,m}$
for the discussion of the initial value dependence of $U_{t,n,m}$ in \eqref{nonlinear_pseudo_Toda1}.
In fact, \eqref{nonlinear_pseudo_Toda1} has a slightly different coprimeness property from that previously discussed: i.e., {\sl the extended coprimeness property}, which we wish to explain in this article.
In our previous studies on the coprimeness property, we have concentrated only on the Laurent monomial factors of the initial variables.
However, to fully understand the integrable and nonintegrable nature of generic nonlinear equations such as \eqref{nonlinear_pseudo_Toda1}, it is necessary for us to also investigate non-monomial factors that give rise to the singularities of the equations, such as $(U_{t,n,m}-1)^{\pm}$. Here is the definition:
\begin{Definition} \label{def12}
Let $\{x_{\n}\}$ be a discrete dynamical system with the independent variable $\n$ and the dependent variable $x$ and let $R$
be the ring of polynomials of the initial variables.
Let $E$ be a subset of $R\setminus\{0\}$.
\begin{itemize}
\item $\{x_{\n}\}$ has the extended Laurent property with respect to the extension factors $E$ if $x_{\n}\in R[f^{-1}\, |\, f\in E]$ for all $\n$.
\item $\{x_{\n}\}$ has the extended coprimeness property with respect to the exclusion factors $E$ if
there exists a positive constant $D$ such that for every pair of iterates $x_{\n}, x_{\n'}$ with a distance $d(\n,\n')\ge D$, they are coprime with each other in the sense of definition \ref{def13} where $\mathcal{A}= R[f^{-1}\, |\, f\in E]$.
\end{itemize}
\end{Definition}
\begin{Definition} \label{def13}
Let $\mathcal{A}$ be a unique factorization domain and \textrm{Q}$(\mathcal{A})$ be its quotient field. Two elements $f,\,g \in$ Q$(\mathcal{A})$ are coprime with each other as rational functions if they have a decomposition $f=f_1/f_2$, $g=g_1/g_2$ where any common factor of arbitrary two elements in $\{f_1,f_2,g_1,g_2\} \subset \mathcal{A}$ is a unit in $\mathcal{A}$.
\end{Definition}
\begin{Remark}
It is crucial to choose an appropriate set $E$.
If $E$ is the set of the monomials of $R$,  definition \ref{def12} is equivalent to the classical Laurent property and the  coprimeness property in definition \ref{def11}.
If $E$ is too large, e.g., $E=R\setminus\{0\}$, the extendend coprimeness property is trivially satisfied regardless of the properties of a given equation.
Therefore we need to take $E$ depending on the singularities of a given equation, so that we can overcome the difficulties resulting from the singularities that trivially arise from the equation itself.
\end{Remark}
This paper is organized as follows:
in section \ref{somossection}, our idea is explained using a simpler case:  an extended Somos-$4$ recurrence and its nonlinear form.
In section \ref{section3}, we shall investigate the extended two-dimensional discrete Toda equation.
Finally in the appendix we give proof of Propositions omitted in the article.
\section{Nonlinear recurrence related to Somos-$4$}\label{somossection}
\subsection{Nonlinear form of extended Somos-$4$ recurrence}
Let us start from the  recurrence relation
\begin{equation}\label{Pseudo-SOMOS4}
x_{n+4}x_n=x_{n+3}^m x_{n+1}^l+x_{n+2}^k.
\end{equation}
Equation \eqref{Pseudo-SOMOS4} is a coprimeness-preserving  extension to the Somos-$4$ recurrence \cite{KMMT2,SOMOS}:
\[
x_{n+4}x_n=x_{n+3}x_{n+1}+x_{n+2}^2,
\]
and is also a reduction from the extended ``bilinear'' form \eqref{pDToda_polinear_eq}.
In this section we study the following nonlinear mapping:
\begin{equation}\label{u-nonlinear}
\frac{(u_{n+4}-1)(u_n-1)}{(u_{n+2}-1)^k}=\frac{u_{n+3}^mu_{n+1}^l}{u_{n+2}^k} \quad (k,l,m \in \Z_{>0}).
\end{equation}
Equation \eqref{u-nonlinear} is given by a reduction  of \eqref{nonlinear_pseudo_Toda1} to a one-dimensional lattice as follows:
\[
u_{2t+n+m}:=U_{t,n,m}, 
\]
with $k_1+k_2 \rightarrow k$, $l_1 \rightarrow l$ and $l_2 \rightarrow m$. 
It is important to note that Equation \eqref{u-nonlinear} is the nonlinear form of the extended Somos-$4$ \eqref{Pseudo-SOMOS4}, which is not trivial and shall be explained in this subsection.
Putting
\begin{equation}\label{def_u}
u_{n+2}:=\frac{x_{n+4}x_n}{x_{n+2}^k}
\end{equation}
and substituting \eqref{def_u} in \eqref{u-nonlinear}, we have
\begin{align*}
(\hat{S}^2-k+\hat{S}^{-2}) \log (u_{n+2}-1)
&=(m\hat{S}-k+l\hat{S}^{-1})\log u_{n+2}\\
&=(m\hat{S}-k+l\hat{S}^{-1})(\hat{S}^2-k+\hat{S}^{-2})\log x_{n+2},
\end{align*}
where $\hat{S}$ is an up-shift operator with respect to $n$. Thus we obatin
\begin{align}
&(\hat{S}^2-k+\hat{S}^{-2})\left(\log(u_{n+2}-1)-(m\hat{S}-k+l\hat{S}^{-1})\log x_{n+2}\right) \notag\\
&=(\hat{S}^2-k+\hat{S}^{-2})\log \left( \frac{x_{n+4}x_n-x_{n+2}^k}{x_{n+3}^mx_{n+1}^l} \right)=0. \label{def_u_next}
\end{align}
Let us introduce a new variable $F_n$ by
\begin{equation}\label{def_f}
F_n:= \frac{x_{n+4}x_n-x_{n+2}^k}{x_{n+3}^mx_{n+1}^l}.
\end{equation}
Then we can rewrite \eqref{def_u_next} as
\begin{equation}\label{f_eq} 
\frac{F_{n+4}F_n}{F_{n+2}^k}=1.
\end{equation}
Equation \eqref{f_eq} is a nine-term recurrence relation for $\{x_n\}$,
whose initial variables are $x_0,x_1,...,x_7$. An iterate $x_n$
is a rational function of these initial variables.
The recurrence \eqref{f_eq} is, in fact, explicitly solvable.
Let us introduce
\begin{equation}\label{f_g_defs}
\begin{array}{ll}
f_0:=F_0=\dfrac{x_4x_0-x_{2}^k}{x_{3}^mx_{1}^l},\qquad &f_1:=F_2=\dfrac{x_6x_2-x_{4}^k}{x_{5}^mx_{3}^l},\\
g_0:=F_1=\dfrac{x_5x_1-x_{3}^k}{x_{4}^mx_{2}^l},\qquad &g_1:=F_3=\dfrac{x_7x_3-x_{5}^k}{x_{6}^mx_{4}^l},
\end{array}
\end{equation}
and the sequence $\{a_n\}$ defined by
\begin{equation}\label{a_seq}
a_{i+1}-ka_{i}+a_{i-1}=0 \ (i=0,1,2,...),\ \  a_{-1}=-1, a_0=0,
\end{equation}
($a_1=1,\,a_2=k,\, a_3=k^2-1,\, a_4=k^3-2k,  ...$).
Then it is easy to prove by induction that
\begin{equation}\label{F_sol}
F_{2i}=\frac{f_1^{a_i}}{f_0^{a_{i-1}}},\quad F_{2i+1}=\frac{g_1^{a_i}}{g_0^{a_{i-1}}}\ \ (i\ge 0).
\end{equation}
Therefore the following pair of recurrences is equivalent to \eqref{f_eq}: 
\begin{subequations}
\begin{align}
x_{2i+4}x_{2i}&=x_{2i+2}^k +\left(\dfrac{f_1^{a_i}}{f_0^{a_{i-1}}}  \right)x_{2i+3}^mx_{2i+1}^l \label{s_eq1},\\
x_{2i+5}x_{2i+1}&=x_{2i+3}^k +\left(\dfrac{g_1^{a_i}}{g_0^{a_{i-1}}}  \right)x_{2i+4}^mx_{2i+2}^l. \label{s_eq2}
\end{align}
\end{subequations}
Note that \eqref{s_eq1} and \eqref{s_eq2} are trivially satisfied for $i=0,1$.
In the case $F_n=1$ for all $n\in\mathbb{Z}$, $k=2$ and  $l=m=1$, we have 
\[
x_{n+4}x_n=x_{n+3}x_{n+1}+x_{n+2}^2,
\]
which is the Somos-$4$ recurrence.
Therefore the equation \eqref{u-nonlinear} is indeed a nonlinear form of the extended Somos-$4$ equation.

Our main results are the extended Laurent property of \eqref{s_eq1} and \eqref{s_eq2}, and the extended coprimeness property of \eqref{u-nonlinear} with respect to some extension/exclusion factors. 
Let us first introduce the former result: proposition \ref{ext-Laurent} which states the extended Laurent  property of \eqref{s_eq1} and \eqref{s_eq2}.
\begin{Proposition}\label{ext-Laurent}
$x_n$ is a Laurent polynomial of the initial data: i.e.,
\[
 x_n \in \mR:=\Z\left[ x_4^\pm, x_5^\pm, x_6^\pm, x_7^\pm, f_0^\pm, f_1^\pm, g_0^\pm, g_1^\pm\right],
\]
\end{Proposition}
which means that the system $\{x_n\}$ has the extended Laurent property with respect to the extension factors
\[
E=\{x_4, x_5, x_6, x_7, x_4x_0-x_2^k, x_5x_1-x_3^k,x_6x_2-x_4^k,x_7x_3-x_5^k\}.
\]
Unlike the Laurent property in the classical sense, $x_n$ is a Laurent polynomial of  $f_0,f_1,g_0,g_1$, which are not the initial variables themselves.
Let us now introduce the latter half of our main result on the nonlinear recurrence \eqref{u-nonlinear}.
\begin{Theorem}\label{conjec-1}
$x_n \in \mR$ are all irreducible and pairwise coprime.
\end{Theorem}
\begin{Theorem}\label{Th2-coprimeness}
The solution $u_n$ of equation \eqref{u-nonlinear} satisfies the following ``coprimeness'' property: 
if  $n \not\equiv n'$ (mod $2$) or $|n-n'|>4$ is satisfied, then two iterates $u_n$ and $u_{n'}$ are co-prime in the following ring $\mR_u$:
\begin{equation}
\mR_u:=\Z\left[ \{u_j^\pm, (u_j-1)^\pm\}_{j=2}^5\right],
\end{equation}
which means that the system $\{x_n\}$ has the extended coprimeness property with respect to the exclusion factors
\[
E=\{u_2, u_3, u_4, u_5, u_2-1, u_3-1, u_4-1, u_5-1\}.
\]
\end{Theorem}
\subsection{Proof of Proposition \ref{ext-Laurent}}
Let us prepare a lemma to facilitate the proof:
\begin{Lemma}\label{lem-1}
Suppose $x_j \in \mR$ for all $j$ with $7 \le j \le n$,
then the four iterates $x_n,\, x_{n-1},\, x_{n-2},\, x_{n-3}$ are mutually co-prime.
\end{Lemma}
\begin{Proof}[Lemma \ref{lem-1}]
It is trivial that $x_7,\,x_6,\,x_5,\,x_4$ are mutually co-prime in $\mR$.
When $n=8$, we have
\[
x_8x_4=x_6^k+\left(\dfrac{f_1^{a_2}}{f_0^{a_1}}  \right)x_{7}^mx_{5}^l.
\]
If we suppose that $x_8$ has a common non-monomial factor with $x_6$,
then this factor should also divide $x_{7}^mx_{5}^l$, which contradicts the
induction hypothesis that $x_7,\,x_6,\,x_5,\,x_4$ are mutually co-prime in $\mR$.
The same argument proves that $x_8$ is co-prime with $x_7$ and $x_5$.
Thus $x_8$ is co-prime with $x_7,\, x_6,\, x_5$.
We can then prove the statement of the lemma \ref{lem-1} by induction.
\end{Proof}
\begin{Proof} [Proposition~\ref{ext-Laurent}]
The statement is trivial if $n\le 9$.
Let us suppose that $x_n\in \mR$ for all $n$ with $n\le 2i+1$,
and prove the case of $n=2i+2$. From equation \eqref{s_eq1}  we have
\[
x_{2i+2}x_{2i-2}=x_{2i}^k +\left(\dfrac{f_1^{a_{i-1}}}{f_0^{a_{i-2}}}  \right)x_{2i+1}^mx_{2i-1}^l.
\]
Let us focus on the factor $x_{2i-2}$ in \eqref{s_eq1} with $i\to i-2$ and write
\begin{equation}
x_{2i}^k=\frac{1}{x_{2i-4}^k}\left[  \left(\frac{f_1^{a_{i-2}}}{f_0^{a_{i-3}}}   \right)^kx_{2i-1}^{km}x_{2i-3}^{kl} +x_{2i-2}  p_{2i} \right], \label{p_2i_eq}
\end{equation}
where $p_j$ is a polynomial in $x_i$ ($i \le j-1$):
\begin{subequations}
\begin{align}
p_{2i}&=k\left(\frac{f_1^{a_{i-2}}}{f_0^{a_{i-3}}} \right)^{k-1}x_{2i-1}^{m(k-1)}x_{2i-3}^{l(k-1)} x_{2i-2}^{k-1}+O(x_{2i-2}^{2k-1})\label{p2i_expansion},\\
p_{2i+1}&=m \left(\frac{g_1^{a_{i-2}}}{g_0^{a_{i-3}}} \right)x_{2i}^mx_{2i-1}^{k(m-1)}x_{2i-3}^{kl}x_{2i-2}^{l-1} + l \left(\frac{g_1^{a_{i-3}}}{g_0^{a_{i-4}}} \right)x_{2i-1}^{km}x_{2i-3}^{k(l-1)}x_{2i-4}^{l}x_{2i-2}^{m-1}\notag \\
&\qquad\qquad \qquad\qquad  +O(x_{2i-2}^{2l-1})+O( x_{2i-2}^{2m-1}). \label{p2i1_expansion}
\end{align}
\end{subequations}
From equation \eqref{s_eq2} with $i\to i-2$ and $i\to i-3$,
\begin{equation}
x_{2i+1}^mx_{2i-1}^l= \frac{1}{x_{2i-3}^mx_{2i-5}^l}\left[x_{2i-1}^{km} x_{2i-3}^{kl} +x_{2i-2}\times p_{2i+1} \right], \label{p_2i1_eq}
\end{equation}
Since $k a_{i-2}=a_{i-1}+a_{i-3}$, we have
\begin{align*}
&\frac{1}{x_{2i-4}^k}\left(\frac{f_1^{a_{i-2}}}{f_0^{a_{i-3}}}   \right)^kx_{2i-1}^{km}x_{2i-3}^{kl} 
+\frac{1}{ x_{2i-3}^mx_{2i-5}^l } \left(\dfrac{f_1^{a_{i-1}}}{f_0^{a_{i-2}}}  \right) x_{2i-1}^{km}x_{2i-3}^{kl} \\
&= \frac{x_{2i-1}^{km}x_{2i-3}^{kl}}{x_{2i-3}^mx_{2i-4}^kx_{2i-5}^l} \left(\dfrac{f_1^{a_{i-1}}}{f_0^{a_{i-2}}}  \right)
\left\{  x_{2i-4}^k+\left(\dfrac{f_1^{a_{i-3}}}{f_0^{a_{i-4}}}  \right) x_{2i-3}^mx_{2i-5}^l   \right\}\\
&=\left\{\frac{x_{2i-1}^{km}x_{2i-3}^{kl}}{x_{2i-3}^mx_{2i-4}^kx_{2i-5}^l} \left(\dfrac{f_1^{a_{i-1}}}{f_0^{a_{i-2}}}  \right)x_{2i-6}\right\}x_{2i-2}.
\end{align*}
Thus,
\[
x_{2i+2}x_{2i-2}=\frac{x_{2i-2}}{x_{2i-3}^mx_{2i-4}^kx_{2i-5}^l} P_{2i+2},
\]
where
\begin{equation}
P_{2i+2}=\left(\frac{f_1^{a_{i-1}}}{f_0^{a_{i-2}}}  \right) x_{2i-1}^{km}x_{2i-3}^{kl}x_{2i-6}
+x_{2i-3}^mx_{2i-5}^lp_{2i}+\left( \frac{f_1^{a_{i-1}}}{f_0^{a_{i-2}}}     \right)x_{2i-4}^kp_{2i+1}, \label{eq_P2i2}
\end{equation}
which is a polynomial of $x_j$ $(0\le j\le 2i)$.
The right hand side is in $\mR$, and at the same time, the four iterates $x_{2i-2},\,x_{2i-3},\,x_{2i-4},\,x_{2i-5}$ must be mutually co-prime from lemma \ref{lem-1}. Therefore we have $x_{2i-3}^mx_{2i-4}^kx_{2i-5}^l \, |\, P_{2i+2}$. 
Thus $x_{2i+2} \in \mR$ is proved.
By using exactly the same argument we obtain $x_{2i+3} \in \mR$.
Thus $x_n \in \mR$ for $n\ge 4$.
\end{Proof}
\subsection{Proof of Theorem \ref{conjec-1}}
Let us fix some notation.
Let us define $f_n:=F_{2n},\,g_n:=F_{2n+1}$.
Note that $f_n$ is expressed as a monic monomial of $f_0,\,f_1$, and $g_n$ as a monic monomial of $g_0,\,g_1$. Let us define the ring $\mR_n$ by
\[
\mR_n:=\Z\left[x_{2n+4}^\pm,x_{2n+5}^\pm,x_{2n+6}^\pm,x_{2n+7}^\pm,f_n^\pm,f_{n+1}^\pm,g_n^\pm,g_{n+1}^\pm\right]\qquad (n=0,1,2,...).
\]
Note that $\mR_0=\mR$.
We prove theorem \ref{conjec-1} by induction.
\begin{itemize}
\item The case of $n=8, 9$:
Note that
\[
x_8=\frac{x_6^k+\left(\dfrac{f_1^{k}}{f_0} \right)x_7^mx_5^l}{x_4},
\]
 is a first order polynomial in $f_0^{-1}$, whose constant term $x_6^k x_4^{-1}$
is co-prime with the coefficient of $f_0^{-1}$. Thus $x_8$ is irreducible, and not a unit element.
Since $x_9$ is a first order polynomial of $g_0^{-1}$, exactly the same argument as in the case of $n=8$ shows that $x_9$ is irreducible and is not a unit.
\item In the case of $n=10,11$:
Let us take $\q=(x_4,x_5,x_6,x_7,f_0,f_1,g_0,g_1)$ and
$\p=(x_6,x_7,x_8,x_9, f_1,f_2,g_1,g_2)$, and use lemma \ref{lemma-common} in the appendix.
Since $\p$ and $\q$ satisfy the conditions in the lemma and the iterate
$x_{10}$ is irreducible in $\mR_1$, we have a factorization
\[
x_{10}=x_8^{i_{10}}x_9^{j_{10}}x_{10}'(\q)\qquad (i_{10},\, j_{10} \in \Z),
\]
where $x'$ is irreducible.
Note that $f_2,g_2$ are units in $\mR [\q^{\pm}]$, and $f_0,f_1$ are units in
$\mR [\p^{\pm}]$. Since $x_8,\,x_9$ are irreducible polynomials and not units, $i_{10}, \, j_{10}\ge 0$. If
\[
x_{10}=\frac{x_8^k+\left(\dfrac{f_1^{a_3}}{f_0^{a_2}} \right)x_9^mx_7^l}{x_6}
\]
 has $x_8$ as a factor, $x_9^m$ must have the factor $x_8$. 
However, $x_8$ and $x_9$ are co-prime with each other, which leads to a contradiction. In the same manner, we conclude that $x_9$ must not be a factor of $x_{10}$.
Therefore $x_{10}$ is irreducible, and is trivially not a unit.
From lemma \ref{lem-1},  $x_{10}$ is co-prime with $x_8$ and $x_9$.
An argument similar to that in the case of $n=10$ shows that we have a factorization
\[
x_{11}=x_8^{i_{11}}x_9^{j_{11}}x_{11}'(\q)\qquad (i_{11},\,j_{11} \in \Z_{\ge 0}).
\]
The rest of the discussion is the same.
\item  In the case of $n=12,13$:
We have a factorization of $x_{12}$ as
\[
x_{12}=x_8^{i_{12}}x_9^{j_{12}}x_{12}'(\q)\qquad (i_{12},\,j_{12} \in \Z_{\ge 0}).
\]
We prove $i_{12}=j_{12}=0$ in the appendix. Thus $x_{12}$ is irreducible.
In the case of $n=13$, we can prove the irreducibility of $x_{13}$ in the same manner.
\item We shall prove that $x_n$ ($n \le 13$) are mutually co-prime.
Let us define $c_n$ as the value of $x_n$ when we substitute $1$ to all the initial data: i.e.,
\[
c_n=x_n \Big|_{\substack{
x_4=x_5=x_6=x_7=1\\
f_0=f_1=g_0=g_1=1}
}.
\]
As we have already proved the irreducibility of $x_n$ for $n\le 13$,
it is sufficient to prove that $c_n\neq c_m$ for every $n\neq m$ $(n,m\le 13)$. We have
\begin{align*}
&c_8=2,\quad c_9=1+2^m,\quad c_{10}=2^k+(1+2^m)^m,\quad c_{11}=c_9^k+c_{10}^m2^l,
\\ 
&c_{12}=\frac{c_{10}^k+c_{11}^mc_9^l}{2},\quad c_{13}=\frac{c_{11}^k+c_{12}^mc_{10}^l}{c_9}.
\end{align*}
It is easy to see that $c_n<c_{n+1}$ since
\[
c_{n+1}=\frac{c_{n-1}^k+c_n^mc_{n-2}^l}{c_{n-3}}>\frac{c_{n-2}^l}{c_{n-3}}c_n^m>c_n^m\ge c_n
\]
for $n\ge 8$.
Therefore $x_n$ are mutually co-prime.
\item In the case of $n=14$: the proof is lengthy and is in the appendix.
\item In the case of $n\ge 15$: If we suppose that $x_{15}$ is not irreducible, we must have $c_n \le c_{13}c_9=3c_{13}$, which is impossible 
when $(k,l,m)\neq (1,1,1)$ since we have already shown that $c_{14}>3c_{13}$.
The case of $(k,l,m)=(1,1,1)$ is also shown to derive a contradiction, since $c_{n}\ge c_{15}=191>3c_{13}=123$ for $n\ge 15$.
We have proved that $x_n \in \mR$ is irreducible, and is mutually co-prime with each other.
\end{itemize}
\subsection{Proof of Theorem \ref{Th2-coprimeness}}
Let us prepare several lemmas.
We rewrite $\xi_n(\u):=x_n(\x(\u))$, where we have used the following notations:
\[
\x:=\{x_4,x_5,x_6,x_7; f_0,f_1,g_0,g_1\}, \u:=\{x_0,x_1,x_2,x_3; u_2,u_3,u_4,u_5\}.
\]
\begin{Lemma}\label{Th2_lem1}
There is a birational irreducible Laurent mapping between the two sets of variables $\x$ and $\u$.
\end{Lemma}
\begin{Proof}
We construct the rational mapping and show that it is invertible by an elemenatry computation. The details are in the appendix.
\end{Proof}
\begin{Lemma}\label{Th2_lem2}
Let us define
\[
\xi_{2i+2}(\u)=:\frac{x_2^{a_{i+1}}}{x_0^{a_{i}}}\tilde{\xi}_{2i+2}(\u),\;\;
\xi_{2i+3}(\u)=:\frac{x_3^{a_{i+1}}}{x_1^{a_{i}}}\tilde{\xi}_{2i+3}(\u).
\]
Then we have
\[
\tilde{\xi}_n(\u) \in \Z\left[\{u_j^\pm, (u_j-1)^\pm\}_{j=2}^5 \right].
\]
\end{Lemma}
\begin{Proof} Since $x_n \in \mR$,
\[
\xi_n(\u) = x_n \in \Z\left[\{u_j^\pm, (u_j-1)^\pm\}_{j=2}^5, \{x_i^\pm\}_{i=0}^3 \right].
\]
Therefore we need to show that $\tilde{\xi}_n(\u)$ is independent of
$x_0,x_1,x_2,x_3$.
We inductively obtain
\begin{align}
x_{2i+2}&=\frac{x_{2i}^k}{x_{2i-2}}u_{2i}=\frac{x_2^{a_{i+1}}}{x_0^{a_{i}}}u_{2i}^{a_1}u_{2i-2}^{a_2}\cdots u_2^{a_{i}}, \label{x-u1}\\
x_{2i+3}&=\frac{x_3^{a_{i+1}}}{x_1^{a_{i}}}u_{2i+1}^{a_1}u_{2i-1}^{a_2}\cdots u_3^{a_{i}}, \label{x-u2}
\end{align}
where each $a_i$ is defined in \eqref{a_seq}.
The term $u_i$ ($i \ge 6$) can be expressed as rational functions of $u_2,...,u_5$ from $\eqref{u-nonlinear}$.
Thus $\tilde{\xi}_n(\u)$ can be expressed using only $u_2,...,u_5$ and this expression is unique.
\end{Proof}
\begin{Proposition}\label{Th2_prop1}
$\tilde{\xi}_n$ is irreducible in $\Z\left[ \{u_j^\pm, (u_j-1)^\pm\}_{j=2}^5 \right]$.
If $n \ne r$, the two terms $\tilde{\xi}_n$ and $\tilde{\xi}_r$ are co-prime.
\end{Proposition}
\begin{Proof}
Recall that $\DIS \xi_n(\u)\in \mR':=\Z\left[\{u_j^\pm, (u_j-1)^\pm\}_{j=2}^5, \{x_i^\pm\}_{i=0}^3 \right]$.
Let us suppose a factorization $\xi_n(\u)=h_1(\u)h_2(\u)$ ($h_1,\, h_2 \in  \mR')$.
From equations \eqref{x-f0} through \eqref{x-g1}, we have
\[
h_i \in \tilde{\mR}:=\Z\left[ x_0^\pm,x_1^\pm,...,x_7^\pm, f_0^\pm,f_1^\pm,g_0^\pm,g_1^\pm \right] \qquad (i=1,2).
\]
Since $x_n(\x)$ is irreducible in $\mR$, it is also irreducible in $\tilde{\mR}$.
Therefore either $h_1$ or $h_2$ is a unit in $\tilde{\mR}$.
We can assume that $h_1$ is a unit and can factorize $h_1$ as
\[
h_1 (\in \tilde{\mR}) =f_0^{\beta_0}g_0^{\gamma_0} f_1^{\beta_1}g_1^{\gamma_1}  \prod_{i=0}^7x_i^{\alpha_i}\ \ (\alpha_i,\beta_i,\gamma_i \in \Z).
\]
By taking the inverse transformation from $\x$ to $\u$, we have that
$h_1 \in \mR'$ is a unit.
Therefore $\xi_n(\u)$ is irreducible in $\mR'$, and thus $\tilde{\xi}_n$
is irreducible in $\Z\left[ \{u_j^\pm, (u_j-1)^\pm\}_{j=2}^5 \right]$.
Next we prove the coprimeness of two arbitrary iterates.
Let us suppose that $\xi_n(\u)$ and $\xi_r(\u)$ ($n \neq r$)
have a common factor $G$ other than monomial ones.
Then, $G$ is a common factor of the iterates in $\tilde{\mR}$, and is not a unit.
Therefore $x_n$ and $x_r$ are not co-prime in $\tilde{\mR}$.
However, from theorem \ref{conjec-1}, they must be co-prime in $\mR$
and thus co-prime in $\tilde{\mR}$, which is a contradiction.
\end{Proof}
\begin{Proof}[Theorem \ref{Th2-coprimeness}]
It is readily proved from
\[
u_n=\frac{x_{n+2}x_{n-2}}{x_n^k}=\frac{\tilde{x}_{n+2} \tilde{x}_{n-2}}{\tilde{x}_n^k},
\]
and from proposition \ref{Th2_prop1}.
\end{Proof}
\section{Nonlinear extended two-dimensional discrete Toda equation} \label{section3}
\subsection{Introduction and main Theorems}
Based on the results on the nonlinear recurrence equation \eqref{u-nonlinear}
in the previous section, we shall study the irreducibility and extended coprimeness properties of the nonlinear  extended two-dimensional discrete Toda equation  \eqref{nonlinear_pseudo_Toda1}.
Let us redefine the independent variables as $\DIS n':=n+\frac{t}{2},\,m':=m-\frac{t}{2}$,
and use the notations
$\n:=(n',m')$，$\n \in \Z^2 \; (t \in 2\Z)$,  $\n \in (\Z+1/2)^2 \;
(t \in 2\Z+1)$，
\[
\e_1=\left(\frac{1}{2},\frac{1}{2}\right),  \quad \e_2 = \left(-\frac{1}{2},\frac{1}{2}\right).
\]
To ease notation, let us abbreviate the prime ${}^{'}$ in $(n',m')$ from here on.
Then equations \eqref{UtoTau} and \eqref{nonlinear_pseudo_Toda1} are equivalent to the following equations:
\begin{equation}\label{def_u_tmn2}
U_{t,\n}=\frac{\tau_{t+1,\n}\tau_{t-1,\n}}{\tau_{t,\n-\e_2}^{k_1}\tau_{t,\n+\e_2}^{k_2}},
\end{equation}
\begin{equation}\label{nonlinear_pseudo_Toda2}
\frac{(U_{t+1,\n}-1)(U_{t-1,\n}-1)}{(U_{t,\n-\e_2}-1)^{k_1}(U_{t,\n+\e_2}-1)^{k_2}}
=\frac{U_{t,\n-\e_1}^{l_1}U_{t,\n+\e_1}^{l_2}}{U_{t,\n-\e_2}^{k_1}U_{t,\n+\e_2}^{k_2}}.
\end{equation}
Let us define the shift operators $\hat{S}_t,\,\hat{S}_1,\,\hat{S}_2$, which defines an up-shift in the directions of $t,\e_1$ and $\e_2$ respectively. Then the equation \eqref{nonlinear_pseudo_Toda2} can be written as:
\begin{align*}
&\left(\hat{S}_t+\hat{S}_t^{-1}-k_1\hat{S}_2^{-1}-k_2\hat{S}_2 \right)\log\left(U_{t,\n}-1 \right) \\
&=\left(l_1\hat{S}_1^{-1} +l_2\hat{S}_1-k_1\hat{S}_2^{-1}-k_2\hat{S}_2   \right) \log U_{t,\n} \\
&=\left(l_1\hat{S}_1^{-1} +l_2\hat{S}_1-k_1\hat{S}_2^{-1}-k_2\hat{S}_2   \right) 
\left(\hat{S}_t+\hat{S}_t^{-1}-k_1\hat{S}_2^{-1}-k_2\hat{S}_2 \right)\log \tau_{t,\n},
\end{align*}
which is equivalent to
\[
\left(\hat{S}_t+\hat{S}_t^{-1}-k_1\hat{S}_2^{-1}-k_2\hat{S}_2 \right)
\log \left[\frac{\tau_{t+1,\n}\tau_{t-1,\n}-\tau_{t,\n-\e_2}^{k_1}\tau_{t,\n+\e_2}^{k_2}}{\tau_{t,\n-\e_1}^{l_1}\tau_{t,\n+\e_1}^{l_2}}       \right]=0.
\]
Thus we obtain a recurrence relation of $\tau_{\n, t}$ as
\begin{equation}\label{Ftnm_equation}
\frac{F_{t+1,\n}F_{t-1,\n}}{F_{t,\n-\e_2}^{k_1}F_{t,\n+\e_2}^{k_2}}=1,
\end{equation}
where
\begin{equation}\label{Ftnm}
F_{t,\n}:=\frac{\tau_{t+2,\n}\tau_{t,\n}-\tau_{t+1,\n-\e_2}^{k_1}\tau_{t+1,\n+\e_2}^{k_2}}{\tau_{t+1,\n-\e_1}^{l_1}\tau_{t+1,\n+\e_1}^{l_2}}.
\end{equation}
Equation \eqref{Ftnm_equation} is nonlinear and is a five-term relation with respect to $t$.
Evolution of \eqref{Ftnm_equation} can be defined by assigning the values of $\tau_{t,\n}$ at $t=0,1,2,3$, and by computing the iterations for $t \ge 4$.
Let us define the sequences $a_{t,\n},\,b_{t,\n} \in \Z$ by the following linear recurrence relation
\begin{equation}
y_{t+1,\n}-k_1y_{t,\n-\e_2}-k_2y_{t,\n+\e_2}+y_{t-1,\n}=0, \label{linear_atnm}
\end{equation}
and the initial data
\begin{align}
&a_{0,\boldsymbol{0}}=1,\quad a_{0,\n\ne\boldsymbol{0}}=0,\quad a_{1,\n}=0,
\label{linear_initial}\\
&b_{1,-\e_2}=1,\quad b_{0,\n}=0,\quad b_{1,\n\ne -\e_2}=0.
\label{linear_initial2}
\end{align}
Then we can explicitly solve $\{F_{t,\n}\}$ as
\begin{equation}
F_{t,\n}=\prod_{\r_0,\r_1}F_{0,\r_0}^{a_{t,\n-\r_0}}F_{1,\r_1-\e_2}^{b_{t,\n-\r_1}},\label{F_tn_values_even}
\end{equation}
where the products are taken over all integers $\r_0$ and $\r_1$.
Therefore the variable $\tau_{t,\n}$ satisfies the following equation:
\begin{equation}
\tau_{t+1,\n}\tau_{t-1,\n}=\tau_{t,\n-\e_2}^{k_1}\tau_{t,\n+\e_2}^{k_2}
+F_{t-1,\n}\tau_{t,\n-\e_1}^{l_1}\tau_{t,\n+\e_1}^{l_2}. \label{simultaneous_tau}
\end{equation}
It follows from \eqref{F_tn_values_even} that $F_{t-1,\n}$ is a monomial of $F_{0,\n}, \, F_{1,\n}$.
Note that \eqref{simultaneous_tau} is trivial for $t=1,2$ for which it coincides with \eqref{Ftnm}.

Let us summarize our main results in this section.
First we prove the extended Laurent and coprimeness properties of equation \eqref{simultaneous_tau}.
\begin{Theorem}\label{Th3}
Let us define a ring of Laurent polynomials (corresponding to $\mR$ in the previous section) as
\begin{equation}\label{Newring}
\mS:=\Z\left[ \{\tau_{2,\n}^\pm,\tau_{3,\n}^\pm\}, \{F_{0,\n}^\pm,F_{1,\n}^\pm \} \right].
\end{equation}
Then, every iterate $\tau_{t,\n}$ $(t \ge 2)$ belongs to $\mS$.
Moreover, $\tau_{t,\n}$ is irreducible and any two iterates are coprime in $\mS$, which means that \eqref{simultaneous_tau} has the extended coprimeness with respect to the exclusion factors
\[
\left\{\tau_{2,\n},\tau_{3,\n}, \tau_{2,\n}\tau_{0,\n}-\tau_{1,\n-\e_2}^{k_1}\tau_{1,\n+\e_2}^{k_2},\tau_{3,\n}\tau_{1,\n}-\tau_{2,\n-\e_2}^{k_1}\tau_{2,\n+\e_2}^{k_2}\right\}_{\n}.
\]
\end{Theorem}
Next we prove our final goal of this paper: theorem of extended coprimeness property of \eqref{nonlinear_pseudo_Toda2}:
\begin{Theorem}\label{mainTh_2Dtoda}
Two iterates $U(t,\n)$ and $U(s,\r)$ of the equation \eqref{nonlinear_pseudo_Toda2} satisfy the following property:
if $|t-s|>2$ or $\r \ne \n,\,\n\pm 2\e_2$, they are co-prime in
\begin{equation}\label{mSU}
\mS_U:=\Z\left[ \left\{ U_{t,\n}^{\pm}, (U_{t,\n}-1)^\pm  \right\}_{t=0,1}  \right],
\end{equation}
which means that \eqref{nonlinear_pseudo_Toda2} has the extended coprimeness property with respect to the exclusion factors $\{U_{0,\n},U_{1,\n},U_{0,\n}-1,U_{1,\n}-1\}_{\n}$.  
\end{Theorem}
\subsection{Proof of Theorem \ref{Th3}}
\begin{Lemma}\label{lem2D_1}
Let us suppose that $\tau_{t,\n} \in \mS$ for every $t \ge t_0$.
Then two iterates $\tau_{t_0,\n}$ and $\tau_{t_0,\r}$ are co-prime if $\n \ne \r$.
\end{Lemma}
\begin{Proof}
From the spatial symmetry of the equation, if we shift the subscripts of the terms
in $\tau_{t_0,\r}$ in the direction of $\n -\r$, we obtain $\tau_{t_0,\n}$.
Thus if $\tau_{t_0,\r}$ is a unit, then we have that $\tau_{t_0,\n}$
is also a unit and is co-prime with $\tau_{t_0,\r}$. Otherwise, there exists a non-unit factor $\tau_{t_0,\r}'$ such that
$\tau_{t_0,\r}=(\mbox{unit})\times \tau_{t_0,\r}'$.
Since $\tau_{t_0,\r}'$ has only a finite number of variables,
$\tau_{t_0,\n}'$ has at least one variable that is not in $\tau_{t_0,\r}'$.
This variable is not a unit element, and thus
$\tau_{t_0,\r}$ and $\tau_{t_0,\n}$ are co-prime.
\end{Proof}

\begin{Lemma}\label{lem2D_2}
Let us suppose that $\tau_{t,\n}, \tau_{t-1,\n'} \in \mS$ for all $\n, \n'$, and suppose that
$\tau_{t,\n}$ is co-prime with four iterates $\tau_{t-1,\n \pm \e_1}, \tau_{t-1,\n \pm \e_2} $.
Then if $\tau_{t+1,\r} \in \mS$, the iterate $\tau_{t+1,\r}$ is also co-prime with $\tau_{t,\r \pm \e_1}, \tau_{t,\r \pm \e_2}$.
\end{Lemma}
\begin{Proof}
  It is immediately obtained from equation \eqref{simultaneous_tau} and lemma \ref{lem2D_1}.
\end{Proof}
\begin{Proposition}\label{propo_2}
We have $\tau_{t,\n} \in \mS$.
\end{Proposition}
\begin{Proof}  The proposition is trivial when $t \le 5$.
For $t\ge 5$, the proof is done by induction. Suppose that  $\tau_{t,\n}\in \mS$ for $t$, then
\[
\tau_{t+1,\n}\tau_{t-1,\n}=\tau_{t,\n-\e_2}^{k_1}\tau_{t,\n+\e_2}^{k_2}
+F_{t-1,\n}\tau_{t,\n-\e_1}^{l_1}\tau_{t,\n+\e_1}^{l_2} \in \mS.
\]
By a direct calculation in the appendix we conclude that there exists a polynomial $P_{t+1,\n} \in \mS$ in $\tau_{t,\n-\e_2}$ and other iterates such that
\[
\tau_{t+1,\n}\tau_{t-1,\n}=\frac{\tau_{t-1,\n}P_{t+1,\n}}{\tau_{t-2,\n-\e_2}^{k_1}\tau_{t-2+\e_2}^{k_2}
\tau_{t-2,\n-\e_1}^{l_1}\tau_{t-2,\n+\e_1}^{l_2}}.
\]
From lemma \ref{lem2D_2},  $\tau_{t-1,\n}$ is co-prime with $\tau_{t-2,\n-\e_2},\, \tau_{t-2+\e_2},\,
\tau_{t-2,\n-\e_1},\, \tau_{t-2,\n+\e_1}$, and satisfies
$\tau_{t+1,\n}\tau_{t-1,\n}\in \mS$.
Therefore $\tau_{t-2,\n-\e_2}^{k_1}\tau_{t-2+\e_2}^{k_2}
\tau_{t-2,\n-\e_1}^{l_1}\tau_{t-2,\n+\e_1}^{l_2}$ divides $P_{t+1,\n}$.
Thus $\tau_{t+1,\n} \in \mS$.
\end{Proof}
\begin{Proof}[Theorem \ref{Th3}]
It is sufficient to prove the irreducibility only for $\tau_{t,\b0}$ ($t \in 2\Z$), $\tau_{t,-\e_2}$ ($t \in 2\Z+1$),
because of the translational symmetries of the equation.
\begin{itemize}
\item In the case of $t=2,3$, the statement is trivial since $\tau_{t,\n}$ is a unit.
\item In the case of $t=4$:
\[
\tau_{4,\boldsymbol{0}}=\frac{\tau_{3,-\e_2}^{k_1}\tau_{3,\e_2}^{k_2}
+F_{2,\b0}\tau_{3,-\e_1}^{l_1}\tau_{3,\e_1}^{l_2}}{\tau_{2,\b0}}. 
\]
Since $\DIS F_{2,\b0}=\frac{F_{1,-\e_2}^{k_1}F_{1,\e_2}^{k_2}}{F_{0,\b0}}$, $\tau_{4,\b0}$ is a first order polynomial of
$F_{0,\b0}^{-1}$, whose coefficient is co-prime with its constant term.
Thus $\tau_{4,\boldsymbol{0}}$ is irreducible and not a unit.
\item In the case of $t=5$, from lemma \ref{lemma-common},  we have
\[
\tau_{5,-\e_2}=\left( \prod_{\n} \tau_{4,\n}^{\alpha_{\n}} \right)\times \tau_{5,-\e_2}'\qquad (\alpha_{\n} \in \Z_{\ge 0}),
\]
where $\tau_{5,-\e_2}'$ is irreducible.
We also have
\[
\tau_{5,-\e_2}=\frac{\tau_{4,-2\e_2}^{k_1}\tau_{4,\b0}^{k_2}
+F_{3,-\e_2}\tau_{4,-\e_1-\e_2}^{l_1}\tau_{4,\e_1-\e_2}^{l_2}}{\tau_{3,-\e_2}}.
\]
Since $\tau_{4,\n}$ and $\tau_{4,\r}$ are mutually co-prime when $\n \ne \r$ from lemma \ref{lem2D_1}, we have $\alpha_{\n}=0\,$ ($\n=\b0,\,-2\e_2,\,\pm\e_1-\e_2$).
The term $\tau_{4,\n}$ is independent of $F_{0,\r}$ ($\r \ne \n$), and is a first order polynomial of $F_{0,\n}^{-1}$ whose constant term is non-zero.
We also have that $F_{3,-\e_2}$ is a monomial of $F_{0,\b0}$ and $F_{0,-2\e_2}$.
Therefore $\tau_{5,-\e_2}$ is independent of all the iterates $F_{0,\n}$ ($\n \ne \b0, \,-2\e_2, \,\pm\e_1-\e_2$).
Thus $\alpha_{\n}=0$ for all $\n \ne \b0,\, -2\e_2, \,\pm\e_1-\e_2$.
We have proved that $\tau_{5,-\e_2}$ is irreducible.
\item In the case of $t=6$, from lemma \ref{lemma-common} we have the following factorization:
\begin{equation}
\tau_{6,\b0}=\left( \prod_{\n} \tau_{4,\n}^{\alpha_{\n}} \right)\times \tau_{6,\b0}'\qquad (\alpha_{\n} \in \Z_{\ge 0}), \label{factorizetau6}
\end{equation}
where $\tau_{6,\b0}'$ is irreducible.
Let us take the initial data as ${}^\forall \n,\; F_{0,\n}=t_{\n}^{-1}$, and take all the other initial data ($F_{1,\n}, \tau_{2,\n}, \tau_{3,\n}$) as $1$.
Then we have
\begin{align*}
F_{2,\n}&=t_{\n},\\
F_{3,\n}&=t_{\n-\e_2}^{k_1}t_{\n+\e_2}^{k_2},\\
F_{4,\n}&=t_{\n-2\e_2}^{k_1^2}t_{\n}^{2k_1k_2-1}t_{\n+2\e_2}^{k_2^2},
\end{align*}
and
\begin{align*}
\tau_{4,\n}&=1+t_{\n},\\
\tau_{5,\n}&=(1+t_{\n-\e_2})^{k_1}(1+t_{\n+\e_2})^{k_2}+t_{\n-\e_2}^{k_1}t_{\n+\e_2}^{k_2}
(1+t_{\n-\e_1})^{l_1}(1+t_{\n+\e_1})^{l_2},\\
\tau_{6,\n}&=\frac{\tau_{5,\n-\e_2}^{k_1}\tau_{5,\n+\e_2}^{k_2}
+t_{\n-2\e_2}^{k_1^2}t_{\n}^{2k_1k_2-1}t_{\n+2\e_2}^{k_2^2}\tau_{5,\n-\e_1}^{l_1}\tau_{5,\n+\e_1}^{l_2}}{1+t_{\n}}.
\end{align*}
Therefore $a_{\n}=0$ for all $\n\not\in \mI_6$ , since the iterate $\tau_{6,\b0}$ depends only on $t_{\n}$ ($\n \in \mI_6$), where
\[
\mI_6:=\left\{\n=j_1\e_1+j_2\e_2\,\big|\,  |j_1|+|j_2|=0,\, 2,\, (j_1,j_2)\in \Z^2  \right\}.
\]
Thus, it is sufficient to prove that $\alpha_{\n}=0$ for $\n \in \mI_6$ one by one.
The proof is elementary and found in the appendix.
From these observations we conclude that $\tau_{6,\b0}$ is irreducible.
\item Preparation for $t\ge 7$:
Let us define $c_t$ as the value of $\tau_{t,\n}$ when we take all the initial data as $1$.
Note that $c_t$ does not depend on $\n$.
If we substitute $1$ for all the initial data in $F_{t,\n}$, we have $F_{t,\n} = 1$, and 
\[
c_3=1,\quad c_4=2,\quad c_{j+1}=\frac{c_j^{k_1+k_2}+c_j^{l_1+l_2}}{c_{j-1}} \; (j \ge 4).
\]
It is easy to see that the $c_j$ are strictly increasing.
Therefore we have shown the following fact: if $\tau_{t,\n}$ and $\tau_{s,\r}$ ($s \ne t$) are both irreducible, then $\tau_{t,\n}$ and $\tau_{s,\r}$ are co-prime.
Also, from lemma \ref{lemma-common},  $\tau_{t,\n}$ and $\tau_{t,\r}$ ($\n \ne \r$) are co-prime if they are both irreducible.
Thus the irreducibility immediately implies the coprimeness.
\item 
In the case of $t=7$:
From lemma \ref{lemma-common},
\begin{align*}
&\tau_{7,-\e_2}=\prod_{\n}\tau_{4,\n}^{\alpha_{\n}}\tau_{7,-\e_2}'=\prod_{\n,\r}\tau_{5,\n}^{\beta_{\n}}\tau_{6,\r}^{\gamma_{\r}} \tau_{7,-\e_2}'',\\
&(\alpha_{\n},\beta_{\n},\gamma_{\r} \in \Z_{\ge 0}),
\end{align*}
where $\tau_{7,-\e_2}',\, \tau_{7,-\e_2}''$ are irreducible.
If we suppose that $\tau_{7,-\e_2}$ is not irreducible, then there exist $\n,\,\r,\, j$ such that
\[
\tau_{7,-\e_2}=(\mbox{unit})\times \tau_{4,\n}\tau_{j,\r} \qquad(j \in \{5,6\}).
\]
Therefore $c_7 \le c_4c_6=2c_6$.
On the other hand,
\[
c_7=\frac{c_6^{k_1+k_2}+c_6^{l_1+l_2}}{c_5}>c_6^{k_1}+c_6^{l_1}\ge 2c_6,
\]
which is a contradiction.
\item In the case of $t \ge 8$, the discussion goes in exactly the same manner.
\end{itemize}
\end{Proof}
\subsection{Proof of Theorem \ref{mainTh_2Dtoda}}
Let us use the following notations:
\begin{align*}
&\btau_t:=\{ \tau_{t,\n} \},\quad \bF_t:=\{ F_{t,\n} \},\quad \bU_t=\{ U_{t,\n} \}, \\
&\bT:=\{\bF_0,\bF_1,\btau_2,\btau_3\},\quad \bW:=\{\btau_0,\btau_1,\bU_1,\bU_2\}
\end{align*}
Then we have $\DIS \mS=\Z\left[\bF_0^\pm,\bF_1^\pm,\btau_2^\pm,\btau_3^\pm   \right]=\Z\left[ \bT \right]$.
Note that from our previous results, all the iterates $\tau_{t,\n} \in \mS$ are irreducible and mutually co-prime.
As before, we have the following lemma whose proof is found in the appendix.
\begin{Lemma}\label{TW_birational}
There is a birational mapping between $\bT$ and $\bW$.
\end{Lemma}
Note that it is immediately shown that the transformations between $\boldsymbol{W}$ and $\boldsymbol{T}$ are made up of irreducible Laurent polynomials.

\begin{Lemma}
Let us define the new variable $\sigma_{t,\n}$ as $\sigma_{t,\n}(\bW):=\tau_{t,\n}(\bT)$.
Then we have the following factorization: 
$\DIS \sigma_{t,\n}=u_{t,\n}\tilde{\sigma}_{t,\n}$,
where $u_{t,\n}$ is a monic Laurent monomial of $\tau_{0,\r}$, $\tau_{1,\r}$, and we have
\[
\tilde{\sigma}_{t,\n} \in \Z\left[\{U_{s,\r}^{\pm},\,(U_{s,\r}-1)^\pm \}_{s=0,1}  \right].
\]
\end{Lemma}
\begin{Proof}
By a direct computation, we have
\begin{align*}
\tau_{2,\n}&=\frac{U_{1,\n}}{\tau_{0,\n}}\left(\tau_{1,\n-\e_2}^{k_1}\tau_{1,\n+\e_2}^{k_2}\right)=\frac{\tau_{1,\n-\e_2}^{k_1}\tau_{1,\n+\e_2}^{k_2}}{\tau_{0,\n}}\cdot U_{1,\n},\\
\tau_{3,\n}&=\frac{U_{2,\n}}{\tau_{1,\n}}\left(\tau_{2,\n-\e_2}^{k_1}\tau_{2,\n+\e_2}^{k_2}\right)\\
&=\frac{U_{2,\n}}{\tau_{1,\n}}\left(\frac{U_{1,\n-\e_2}}{\tau_{0,\n-\e_2}}\right)^{k_1}
\left(\frac{U_{1,\n+\e_2}}{\tau_{0,\n+\e_2}}\right)^{k_2}\left(\tau_{1,\n-2\e_2}^{k_1^2}\tau_{1,\n}^{2k_1k_2}\tau_{1,\n+2\e_2}^{k_2^2}  \right)\\
&=\frac{\tau_{1,\n-2\e_2}^{k_1^2}\tau_{1,\n}^{2k_1k_2-1}\tau_{1,\n+2\e_2}^{k_2^2}   }{ \tau_{0,\n-\e_2}^{k_1}  \tau_{0,\n+\e_2}^{k_2}  }\cdot U_{2,\n}U_{1,\n-\e_2}^{k_1}U_{1,\n+\e_2}^{k_2},\\
&\hspace{2mm}\vdots \\
\tau_{t,\n}&=(\mbox{monic Laurent monomial of}\ \tau_{0,\r}, \tau_{1,\r} )\\
&\qquad \times
(\mbox{monic monomial of}\ U_{1,\r},...,U_{t-1,\r}).
\end{align*}
From equation \eqref{nonlinear_pseudo_Toda2}, $U_{s,\n}$ $(s=3,4,...)$ can be expressed using $U_{1,\r},\,U_{2,\r}$.
Therefore the former half of $\tau_{t,\n}$ is a monic Laurent monomial of $\tau_{0,\r}$, $\tau_{1,\r}$,
and the latter half  shall be denoted as $\tilde{\sigma}_{t,\n}$, which is in $\Q(\bU_0,\bU_1)$.
On the other hand, from $\tau_{t,\n}\in \mS$ and the transformations \eqref{trans1} -- \eqref{trans4} we have
\[
\sigma_{t,\n} \in \Z\left[\btau_0^\pm,\btau_1^\pm, \{U_{s,\r}^{\pm},\,(U_{s,\r}-1)^\pm \}_{s=0,1} \right].
\]
Therefore, from the uniqueness of the factorization, we conclude that 
\[
\tilde{\sigma}_{t,\n} \in \Z\left[\{U_{s,\r}^{\pm},\,(U_{s,\r}-1)^\pm \}_{s=0,1}  \right].
\]
\end{Proof}
\begin{Proposition}\label{main_prop}
The term $\tilde{\sigma}_{t,\n}$ is irreducible in $\DIS \Z\left[\{U_{s,\r}^{\pm},\,(U_{s,\r}-1)^\pm \}_{s=0,1}  \right]
$.
For $(t,\n) \ne (s,\r)$, $\tilde{\sigma}_{t,\n}$ and $\tilde{\sigma}_{s,\r}$ are co-prime.
\end{Proposition}
\begin{Proof} We use an argument similar to that in the proof of proposition \ref{Th2_prop1}. We have
\[
\sigma_{t,\n} \in \tilde{\mS}:=\Z\left[\btau_0^\pm,\btau_1^\pm, \{U_{s,\r}^{\pm},\,(U_{s,\r}-1)^\pm \}_{s=0,1} \right].
\]
Let us suppose that we can factor $\sigma_{t,\n}$ as $\sigma_{t,\n}=h_1(\bW)h_2(\bW)$ $(h_1,\,h_2 \in \tilde{\mS})$.
From equations \eqref{trans1}--\eqref{trans4} we can reformulate $(U_{s,\r}-1)$, to obtain
\[
h_1h_2 \in \mS':=\Z\left[\{\btau_i^\pm\}_{i=0}^3,\,\bF_0^\pm ,\bF_1^\pm  \right].
\]
On the other hand, since $\tau_{t,\n} \in \mS$ is irreducible in $\mS'$ and 
we can assume that $h_1$ is a unit in $\mS'$.
Therefore $h_1$ can be expressed as
\[
h_1=\prod_{i=0}^1\prod_{\n}F_{i,\n}^{\alpha_{i,\n}}\prod_{j=0}^3\prod_{\r}\tau_{j,\r}^{\beta_{j,\r}}\qquad (\alpha_{i,\n}, \beta_{j,\r} \in \Z),
\]
which implies that $h_1$ is a unit also in $\tilde{\mS}$.
Thus $\sigma_{t,\n}$ is irreducible in $\tilde{\mS}$.
From the uniqueness of the factorization, the iterate $\tilde{\sigma}_{t,\n}$ is irreducible in $\DIS \Z\left[\{U_{s,\r}^{\pm},\,(U_{s,\r}-1)^\pm \}_{s=0,1}  \right]$.
Finally we prove the coprimeness.
Let us suppose that $\sigma_{t,\n}$ and $\sigma_{s,\r}$  are not co-prime.
Then they must have a (non-unit) common factor $H$ in $\tilde{\mS}$. Then $H$ is not a unit in $\mS'$, and $\tau_{t,\n}$ and $\tau_{s,\r}$ are not co-prime in $\mS'$.
This conclusion contradicts the outcome of theorem \ref{Th3} that $\tau_{t,\n}$ and $\tau_{s,\r}$ are co-prime in $\mS$.
\end{Proof}
\hbreak
\begin{Proof}[Theorem \ref{mainTh_2Dtoda}] 
From equation \eqref{nonlinear_pseudo_Toda2}, we have that $U_{t,\n}\in \Q(\bU)$. Thus
\[
U_{t,\n}=\frac{\sigma_{t+1,\n}\sigma_{t-1,\n}}{\sigma_{t,\n-\e_2}^{k_1}\sigma_{t,\n+\e_2}^{k_2}}
=\frac{\tilde{\sigma}_{t+1,\n}\tilde{\sigma}_{t-1,\n}}{\tilde{\sigma}_{t,\n-\e_2}^{k_1}\tilde{\sigma}_{t,\n+\e_2}^{k_2}}
\]
Therefore we obtain theorem \ref{mainTh_2Dtoda} from proposition \ref{main_prop}.
\end{Proof}

\section{Concluding remarks}
In this article, we introduced the extended coprimeness property and the extended Laurent property for discrete dynamical systems and investigated their effectiveness in the study of singularity structures.
As examples, we proved that the nonlinear extended discrete $2$-dimensional Toda lattice equations \eqref{nonlinear_pseudo_Toda1} all have the coprimeness property in $\Z\left[\{U_{s,\r}^{\pm},\,(U_{s,\r}-1)^\pm \}_{s=0,1}  \right]$
 for general $k_1,k_2,l_1$ and $l_2$: i.e., it has the extended coprimeness with respect to the exclusion factors $E=\{U_{0,\r},(U_{0,\r}-1),U_{1,\r},(U_{1,\r}-1)\}$. 
 
The nonlinear recurrence \eqref{u-nonlinear} corresponding to the extended Somos-$4$,  which is obtained as a reduction of \eqref{nonlinear_pseudo_Toda1},  also possesses the coprimeness property in the ring $\DIS \Z\left[ \{u_j^\pm, (u_j-1)^\pm\}_{j=2}^5\right]$:
i.e., with the exclusion factors
$\{u_j, u_j-1 \}_{j=2}^5$. 
Since the coprimeness property is an algebraic analogue of singularity confinement \cite{SC},
the reason why the ring $\DIS \Z\left[ \{u_j^\pm, (u_j-1)^\pm\}_{j=2}^5\right]$ (or the the exclusion factors
$\{u_j, u_j-1 \}_{j=2}^5$) appears in the property is that the \lq\lq singularities\rq\rq of \eqref{u-nonlinear} are not only at $u_j=0$  but also at $u_j=1$  $(j=2,3,4,5)$.

So far we have constructed several higher dimensional nonlinear equations which have the coprimeness property\cite{KMT,KKMT}. A natural question is whether there is any systematic way of constructing such equations. 
For discrete Painlev\'{e} equations, which were first introduced as second order rational mappings with the singularity confinement property, Sakai has given a geometric construction of the so called space of initial conditions and has completed the classification of the discrete Painlev\'{e} equations\cite{RGH,Sakai}.
We expect some geometric interpretation of the coprimeness property with which a
systematic construction and classification of coprimeness preserving nonlinear equations becomes possible.
This is one of the problems we wish to address in future works.
An investigation of the continuous limits of these discrete equations is also a future problem. 

\section*{Acknowledgement}
The authors are grateful to Prof. R. Willox for useful comments.
The present work is partially supported by KAKENHI Grant Numbers 15H06128, 16H06711, and JP21340034.

\appendix

\section{Lemma on the change of variables}
Let us prepare a lemma on the factorization of Laurent polynomials with respect to a change of variables, which has been introduced in \cite{KMMT2}
\begin{Lemma}[\cite{KMMT2}] \label{lemma-common}
Let $\mK$ be a unique factorization domain (UFD).
Suppose that we have a bijective mapping between $\p=\{p_1,p_2,...,p_N\}$ and $\q=\{q_1,q_2,...,q_N\}$ such that $\q \in \mK[\p^\pm]$, $\p \in \mK[\q^\pm]$,
and that each $q_i$ in $\q$ is irreducible in $\mK[\p^{\pm}]$.
Let us take a Laurent polynomial $f(\q) \in \mK[\q^\pm]$, which
is irreducible in $\mK[\p^\pm]:$ i.e., $\tilde{f}(\p):=f(\q(\p))\in\mK[\p^\pm]$ is irreducible.
Then $f(\q)$ admits the following factorization in $\mK[\q^\pm]:$
\[
f(\q)=\p(\q)^{\boldsymbol{\alpha}}f'(\q) \qquad (\boldsymbol{\alpha}\in \Z^N,\, f'\, \mbox{is irreducible}).
\]
\end{Lemma}
Sketch of the proof:  The conditions show that the Laurent polynomial $f$ is an irreducible element in $\mK[\q^\pm, \p^\pm]$.

\section{Proof of several Propositions}
\subsection{Case of $t=12,14$ in proof of Theorem \ref{conjec-1}}
In the case of $n=12$ we have a factorization of $x_{12}$ as
\[
x_{12}=x_8^{i_{12}}x_9^{j_{12}}x_{12}'(\q)\qquad (i_{12},\,j_{12} \in \Z_{\ge 0}).
\]
Let us prove $i_{12}=0$.
Let  $f_1=t$ and take all the other initial variables as $1$. Then we have
\begin{align*}
x_8&=1+t^{a_2}=1+t^k, x_9=1+x_8^m, \\
x_{10}&=x_8^k+t^{k^2-1}x_9^m, x_{11}=x_9^k+x_{10}^mx_8^l.
\end{align*}
If we substitute $t=\ex^{\sqrt{-1}\pi/k}$ then
\[
x_8=0,\quad x_9=1,\quad x_{10}=t^{k^2-1},\quad x_{11}=1.
\]
Since we have
\[
p_{10}=\delta_{k,1}, \quad p_{11}=m t^{m(k^2-1)}\delta_{l,1}+l\delta_{m,1},
\]
we can obtain $P_{12}$ as
\begin{align*}
P_{12}&=t^{a_4}+\delta_{k,1}+t^{a_4}\left(m t^{m(k^2-1)}\delta_{l,1}+l\delta_{m,1} \right)\\
&=\delta_{k,1}+(-1)^k \left\{ 1+  m t^{m(k^2-1)}\delta_{l,1}+l\delta_{m,1}         \right\}.
\end{align*}
Here we have used the fact that $t^{a_4}=(-1)^{k^2}=(-1)^k$ since $a_4=k(k^2-2)$.
Therefore we have $P_{12} \ne 0$ if $k\ne 1$.
We conclude that $P_{12}=0$
if and only if $k=1$ and $m \ge 2,\, l \ge 2$.
Thus, except for the cases of $k=1, \, l\ge 2,\,m \ge 2$, the iterate $x_{12}$ cannot have $x_8$ as a factor.
Let us study the case of $k=1,\, l\ge 2,\,m \ge 2$.
We have
\[
x_9 \equiv \frac{x_7}{x_5},\quad x_{11} \equiv \frac{x_9}{x_7}\equiv \frac{1}{x_5},\quad p_{10}\equiv 1, \quad p_{11} \equiv 0,
\]
where $A\equiv B$ is considered as modulo the factor $x_8$.
We also have $ a_3=0,\,a_4=-1$. Thus
\begin{align*}
P_{12} &\equiv \left(\frac{f_1^{a_{4}}}{f_0^{a_{3}}}  \right) x_{9}^{m}x_{7}^{l}x_{4}
+x_{7}^mx_{5}^l\equiv \frac{x_7^{m+l}x_4}{x_5^mf_1}+x_7^mx_5^l \not\equiv 0.
\end{align*}
Therefore $P_{12}$ cannot have $x_8$ as a factor, and thus $i_{12}=0$.
We can prove that $j_{12}=0$ in a similar manner.
Next in the case of $n = 14$, using the lemma \ref{lemma-common}, we have the following two factorizations
\[
x_{14}=x_8^{r_1}x_9^{r_2}x_{14}'=x_{10}^{r_3}x_{11}^{r_4}x_{12}^{r_5}x_{13}^{r_6}x_{14}'',
\]
where $x_{14}',\,x_{14}''$ are irreducible in $\mR$. 
Let us suppose that $x_{14}$ is {\bf not} irreducible. Then the only possible factorization is
\[
x_{14}=\alpha x_ix_j\qquad (i \in \{8,9\},\ j \in \{10,11,12,13\}),
\]
where $\alpha$ is a unit in $\mR$.
Therefore $c_{14}=c_i c_j \le c_9 c_{13}$.
On the other hand, since we have
\[
c_{14}=\frac{c_{12}^k+c_{13}^mc_{11}^l}{c_{10}},
\]
\[
c_{14}>c_{13}c_{11}>c_{13}c_9,
\]
in the case of $m \ge 2$ or $l \ge 2$, which contradicts $c_{14}\le c_9 c_{13}$.
In the case of $m=l=1$, we have
\[
c_8=2,\quad c_9=3,\quad c_{10}=2^k+3,\quad c_{11}=2^{k+1}+3^k +6,
\]
and thus, when $k\ge 3$,
\[
c_{14}>\frac{c_{13}c_{11}}{c_{10}}=c_{13}\frac{2^{k+1}+3^k+6}{2^k+3}>3c_{13}=c_{13}c_{9},
\]
which also leads to a contradiction.
The remaining cases are $(k,l,m)=(2,1,1)$ and $(1,1,1)$.
If $(k,l,m)=(2,1,1)$ we can directly confirm that $c_{14}=1529>3c_{13}=942$.
If $(k,l,m)=(1,1,1)$, we have $c_{14}=111=3\times 37$, which cannot be expressed as 
$c_i c_j$, $(8\le i\le 9, 10\le j\le 13)$.
We have proved that $x_{14}$ is irreducible.
\subsection{Proof of Lemma \ref{Th2_lem1}}
From the definition of $u_i$ \eqref{def_u},
\begin{equation}\label{x-u-rel}
x_4=\frac{x_2^k}{x_0}u_2, x_5=\frac{x_3^k}{x_1}u_3,
x_6=\frac{x_4^k}{x_2}u_4=\frac{x_2^{k^2-1}}{x_0^k}u_2^ku_4,
x_7=\frac{x_5^k}{x_3}u_5=\frac{x_3^{k^2-1}}{x_1^k}u_3^ku_5.
\end{equation}
From \eqref{f_g_defs} we have
\begin{subequations}
\begin{align}
f_0&=\frac{x_4x_0-x_2^k}{x_3^mx_1^l}=\frac{x_2^k(u_2-1)}{x_3^mx_1^l},\label{x-f0}\\
f_1&=\frac{x_6x_2-x_4^k}{x_5^mx_3^l}=\frac{x_4^k(u_4-1)}{x_5^mx_3^l}
=\frac{x_2^{k^2}x_1^m u_2^k(u_4-1)}{u_3^mx_3^{km+l}x_0^k}\label{x-f1},\\
g_0&=\frac{x_5x_1-x_3^k}{x_4^mx_2^l}=\frac{x_3^kx_0^m(u_3-1)}{x_2^{mk+l}u_2^m}\label{x-g0},\\
g_1&=\frac{x_7x_3-x_5^k}{x_6^mx_4^l}=\frac{x_5^kx_2^m(u_5-1)}{x_4^{mk+l}u_4^m}
=\frac{x_3^{k^2}x_2^mx_0^{mk+l}u_3^k(u_5-1)}{x_1^kx_2^{k(mk+l)}u_2^{mk+l}u_4^m}\label{x-g1}.
\end{align}
\end{subequations}
The inverse mapping is
\[
x_3=\frac{x_4^lx_6^m g_1+x_5^k}{x_7},\ x_2=\frac{x_5^mx_3^lf_1+x_4^k}{x_6},\  
x_1=\frac{x_2^lx_4^m g_0+x_3^k}{x_5},\  x_0=\frac{x_3^mx_1^lf_0+x_2^k}{x_4},
\]
and
\[
u_j=\frac{x_{j+2}x_{j-2}}{x_j^k}\  (j=2,3,4,5).
\]
It is easy to see that $\u$ is expressed as irreducible Laurent polynomials of $\x$, and vice versa.
\subsection{Direct calculation in Proposition \ref{propo_2}}
By a direct calculation we have
\begin{align*}
\tau_{t+1,\n}\tau_{t-1,\n}&=\frac{1}{\tau_{t-2,\n-\e_2}^{k_1}\tau_{t-2,\n+\e_2}^{k_2}
\tau_{t-2,\n-\e_1}^{l_1}\tau_{t-2,\n+\e_1}^{l_2}}\\
&\times  \Big[   \tau_{t-2,\n-\e_1}^{l_1}\tau_{t-2,\n+\e_1}^{l_2}F_{t-2,\n-\e_2}^{k_1}\tau_{t-1,\n-\e_1-\e_2}^{l_1k_1} \\
&\qquad \cdot \tau_{t-1,\n+\e_1-\e_2}^{l_2k_1}   
F_{t-2,\n+\e_2}^{k_2}\tau_{t-1,\n-\e_1+\e_2}^{l_1k_2} \tau_{t-1,\n+\e_1+\e_2}^{l_2k_2}    \\
&\quad  +\tau_{t-2,\n-\e_2}^{k_1}\tau_{t-2,\n+\e_2}^{k_2}F_{t-1,\n}
\tau_{t-1,\n-\e_2-\e_1}^{k_1l_1}\tau_{t-1,\n+\e_2-\e_1}^{k_2l_1}     
\tau_{t-1,\n-\e_2+\e_1}^{k_1l_2}\tau_{t-1,\n+\e_2+\e_1}^{k_2l_2} \\
&\qquad +
\tau_{t-1,\n}\times\left( \mbox{polynomials of}\ \tau_{t-1,\n-2\e_2}  \right)             
 \Big].
\end{align*}
We further compute the first two terms in the square brackets above and obtain
\begin{align*}
&\tau_{t-1,\n-\e_1-\e_2}^{l_1k_1} \tau_{t-1,\n+\e_1-\e_2}^{l_2k_1}\tau_{t-1,\n-\e_1+\e_2}^{l_1k_2} \tau_{t-1,\n+\e_1+\e_2}^{l_2k_2}       \\
& \quad \times \left(\tau_{t-2,\n-\e_1}^{l_1}\tau_{t-2,\n+\e_1}^{l_2}F_{t-2,\n-\e_2}^{k_1}F_{t-2,\n+\e_2}^{k_2}+  \tau_{t-2,\n-\e_2}^{k_1}\tau_{t-2,\n+\e_2}^{k_2}F_{t-1,\n}                           \right)  \\
&=\tau_{t-1,\n-\e_1-\e_2}^{l_1k_1} \tau_{t-1,\n+\e_1-\e_2}^{l_2k_1}\tau_{t-1,\n-\e_1+\e_2}^{l_1k_2} \tau_{t-1,\n+\e_1+\e_2}^{l_2k_2} F_{t-1,\n} \\
&\quad \times\left( \tau_{t-2,\n-\e_2}^{k_1}\tau_{t-2,\n+\e_2}^{k_2}+F_{t-3,\n}\tau_{t-2,\n-\e_1}^{l_1}\tau_{t-2,\n+\e_2}^{l_2}   \right)\\
&=\tau_{t-1,\n-\e_1-\e_2}^{l_1k_1} \tau_{t-1,\n+\e_1-\e_2}^{l_2k_1}\tau_{t-1,\n-\e_1+\e_2}^{l_1k_2} \tau_{t-1,\n+\e_1+\e_2}^{l_2k_2} F_{t-1,\n}\tau_{t-1,\n}\tau_{t-3,\n}.
\end{align*}
\subsection{Case of $t=6$ in proof of Theorem \ref{Th3}}
Let us prove that $\alpha_{\n}=0$ for $\n \in \mI_6$.
\begin{description}
\item[(i)] Substituting $t_{2\e_1}=-1$ and $t_{\n}=1$ $(\n\neq 2\e_1)$, we have $\tau_{4,2\e_1}=0$ and
\[
\tau_{5,\pm\e_2}=2^{k_1+k_2}+2^{l_1+l_2}=:c_5,
\quad
\tau_{5,\e_1}=2^{k_1+k_2},\quad \tau_{5,-\e_1}=c_5.
\]
Therefore we have $\tau_{6,\b0}>0$. 
It follows from the factorization \eqref{factorizetau6} that $\alpha_{2\e_1}=0$.
From the symmetry of the equation we also have $\alpha_{-2\e_1}=0$.
\item[(ii)] Substituting $t_{2\e_2}=-1$ and $t_{\n}=1$ $(\n\neq 2\e_2)$, we have $\tau_{4,2\e_2}=0$ and
\[
\tau_{5,\e_2}=(-1)^{k_2}2^{l_1+l_2},\quad \tau_{5,-\e_2}=c_5,
\quad \tau_{5,\pm \e_1}=c_5.
\]
Since $k_2^2 \equiv k_2$ (mod $2$)  we have
\[
\tau_{6,\b0}=(-1)^{k_2^2}\frac{ 2^{(l_1+l_2)k_2}c_5^{k_1}+c_5^{l_1+l_2}   }{2}\ne 0.
\]
Therefore $\alpha_{2\e_2}=0$. We also have $\alpha_{-2\e_2}=0$ in the same manner.
\item[(iii)] Substituting $t_{\e_1+\e_2}=-1$ and $t_{\n}=0$ $(\n\neq \e_1+\e_2)$ we have
\[
\tau_{5,\e_2}=1, \tau_{5,-\e_2}=1,\tau_{5,\e_1}=0, \tau_{5,-\e_1}=1.
\]
Therefore
$\tau_{6,\b0}=1$ and thus $\alpha_{\e_1+\e_2}=0$. We also have $\alpha_{-\e_1-\e_2}=0$, $\alpha_{\e_1-\e_2}=0$, and, $\alpha_{-\e_1+\e_2}=0$.
\item[(iv)] Substituting $t_{\b0}=-1$ and $t_{\n}=1$ $(\n\neq \b0)$ we have
\[
F_{3,\e_2}=(-1)^{k_1}, F_{3,-\e_2}=(-1)^{k_2}, F_{3,\n}=1 \ (\n \neq \pm \e_2), F_{4,\b0}=-1,
\]
and $P_{6,\b0}\neq 0$, which is proved below. Thus we have $\alpha_{\b0}=0$.
\end{description}
Let us calculate $P_{t+1,\n}$:
\[
\tau_{t+1,\n}=\frac{\tau_{t,\n-\e_2}^{k_1}\tau_{t,\n+\e_2}^{k_2}+F_{t-1,\n}\tau_{t,\n-\e_1}^{l_1}\tau_{t,\n+\e_1}^{l_2}}{\tau_{t-1,\n}},
\]
\begin{align*}
\tau_{t,\n-\e_2}^{k_1}&=\frac{1}{\tau_{t-2,\n-\e_2}^{k_1}}
\left( F_{t-2,\n-\e_2}^{k_1}\tau_{t-1,\n-\e_1-\e_2}^{k_1l_1}\tau_{t-1,\n+\e_1-\e_2}^{k_1l_2}\right. \\
&\left. \quad +k_1F_{t-2,\n-\e_2}^{k_1-1}\tau_{t-1,\n-\e_1-\e_2}^{(k_1-1)l_1}\tau_{t-1,\n+\e_1-\e_2}^{(k_1-1)l_2}\tau_{t-1,\n-2\e_2}^{k_1}\tau_{t-1,\n}^{k_2}+O( \tau_{t-1,\n}^{2k_2})   \right),\\
\tau_{t,\n+\e_2}^{k_2}&=\frac{1}{\tau_{t-2,\n+\e_2}^{k_2}}
\left( F_{t-2,\n+\e_2}^{k_2}\tau_{t-1,\n-\e_1+\e_2}^{k_2l_1}\tau_{t-1,\n+\e_1+\e_2}^{k_2l_2}\right. \\
&\left. \quad +k_2F_{t-2,\n+\e_2}^{k_2-1}\tau_{t-1,\n-\e_1+\e_2}^{(k_2-1)l_1}\tau_{t-1,\n+\e_1+\e_2}^{(k_2-1)l_2}\tau_{t-1,\n}^{k_1}\tau_{t-1,\n+2\e_2}^{k_2}+O( \tau_{t-1,\n}^{2k_1})   \right),\\
\tau_{t,\n-\e_1}^{l_1}&=\frac{1}{\tau_{t-2,\n-\e_1}^{l_1}}
\left( \tau_{t-1,\n-\e_1-\e_2}^{l_1k_1}\tau_{t-1,\n-\e_1+\e_2}^{l_1k_2}                         \right. \\
&\left. \quad + l_1\tau_{t-1,\n-\e_1-\e_2}^{k_1(l_1-1)}\tau_{t-1,\n+\e_2-\e_1}^{k_2(l_1-1)}F_{t-2,\n-\e_1}\tau_{t-1,\n-2\e_1}^{l_1}\tau_{t-1,\n}^{l_2}+O(\tau_{t-1,\n}^{2l_2})                         \right),\\
\tau_{t,\n+\e_1}^{l_2}&=\frac{1}{\tau_{t-2,\n+\e_1}^{l_2}}
\left( \tau_{t-1,\n+\e_1-\e_2}^{l_2k_1}\tau_{t-1,\n+\e_1+\e_2}^{l_2k_2}                         \right. \\
&\left. \quad + l_2\tau_{t-1,\n+\e_1-\e_2}^{k_1(l_2-1)}\tau_{t-1,\n+\e_2+\e_1}^{k_2(l_2-1)}F_{t-2,\n+\e_1}\tau_{t-1,\n}^{l_1}\tau_{t-1,\n+2\e_1}^{l_2}+O(\tau_{t-1,\n}^{2l_1})                         \right).
\end{align*}
Therefore we have
\begin{align}
&P_{t+1,\n}=F_{t-1,\n}\tau_{t-1,\n-\e_1-\e_2}^{l_1k_1} \tau_{t-1,\n+\e_1-\e_2}^{l_2k_1}\tau_{t-1,\n-\e_1+\e_2}^{l_1k_2} \tau_{t-1,\n+\e_1+\e_2}^{l_2k_2} \tau_{t-3,\n}\notag\\
&+\tau_{t-2,\n-\e_1}^{l_1}\tau_{t-2,\n+\e_1}^{l_2}\Big(   
k_2F_{t-2,\n-\e_2}^{k_1}F_{t-2,\n+\e_2}^{k_2-1}\notag \\
&\quad \times \tau_{t-1,\n-\e_1-\e_2}^{k_1l_1}\tau_{t-1,\n+\e_1-\e_2}^{k_1l_2}
\tau_{t-1,\n-\e_1+\e_2}^{(k_2-1)l_1}\tau_{t-1,\n+\e_1+\e_2}^{(k_2-1)l_2}\tau_{t-1,\n}^{k_1-1}\tau_{t-1,\n+2\e_2}^{k_2}   \notag\\
&\left. +k_1F_{t-2,\n+\e_2}^{k_2}F_{t-2,\n-\e_2}^{k_1-1}\tau_{t-1,\n-\e_1+\e_2}^{k_2l_1}\tau_{t-1,\n+\e_1+\e_2}^{k_2l_2}
\tau_{t-1,\n-\e_1-\e_2}^{(k_1-1)l_1}\tau_{t-1,\n+\e_1-\e_2}^{(k_1-1)l_2}\tau_{t-1,\n-2\e_2}^{k_1}\tau_{t-1,\n}^{k_2-1}
\right)\notag\\
&+\tau_{t-2,\n-\e_2}^{k_1}\tau_{t-2,\n+\e_2}^{k_2}F_{t-1,\n}\Big(  
 l_2F_{t-2,\n+\e_1}\tau_{t-1,\n-\e_1-\e_2}^{l_1k_1}\notag\\
&\quad \times\tau_{t-1,\n-\e_1+\e_2}^{l_1k_2}                        
\tau_{t-1,\n+\e_1-\e_2}^{k_1(l_2-1)}\tau_{t-1,\n+\e_2+\e_1}^{k_2(l_2-1)}\tau_{t-1,\n}^{l_1-1}\tau_{t-1,\n+2\e_1}^{l_2} \notag\\
&\left. +l_1F_{t-2,\n-\e_1}\tau_{t-1,\n+\e_1-\e_2}^{l_2k_1}\tau_{t-1,\n+\e_1+\e_2}^{l_2k_2}                        
\tau_{t-1,\n-\e_1-\e_2}^{k_1(l_1-1)}\tau_{t-1,\n+\e_2-\e_1}^{k_2(l_1-1)}\tau_{t-1,\n-2\e_1}^{l_1}\tau_{t-1,\n}^{l_2-1}
 \right)\notag \\
 &+O(\tau_{t-1,\n}).
\label{P_t1_value}
\end{align}
From equation \eqref{P_t1_value},
\begin{align*}
&P_{6,\b0}=-2^{(k_1+k_2)(l_1+l_2)}+k_2\delta_{k_1,1}(-1)^{k_1k_2}(-1)^{k_1(k_2-1)}2^{(k_1+k_2-1)(l_1+l_2)+k_2}\\
&+k_1\delta_{k_2,1}(-1)^{k_1k_2}(-1)^{k_2(k_1-1)}2^{(k_1+k_2-1)(l_1+l_2)+k_1}\\
&+(-l_2)\delta_{l_1,1}2^{(k_1+k_2)(l_1+l_2-1)+l_2}+(-l_1)\delta_{l_2,1}2^{(k_1+k_2)(l_1+l_2-1)+l_1}
\end{align*}
\begin{align*}
&=-2^{(k_1+k_2)(l_1+l_2)}+(-k_2)\delta_{k_1,1}2^{k_2(l_1+l_2+1)}
+(-k_1)\delta_{k_2,1}2^{k_1(l_1+l_2+1)}\\
&+(-l_2)\delta_{l_1,1}2^{l_2(k_1+k_2+1)}+(-l_1)\delta_{l_2,1}2^{l_1(k_1+k_2+1)} \neq 0.
\end{align*}

\subsection{Proof of Lemma \ref{TW_birational}}
First we compute $\bW \, \rightarrow \, \bT$. 
\begin{subequations}
\begin{align}
\tau_{2,\n}&=\frac{1}{\tau_{0,\n}}U_{1,\n}\tau_{1,\n-\e_2}^{k_1}\tau_{1,\n+\e_2}^{k_2},\label{trans1}\\
\tau_{3,\n}&=\frac{U_{2,\n}U_{1,\n-\e_2}^{k_1}U_{1,\n+\e_2}^{k_2} \tau_{1,\n-2\e_2}^{k_1^2}\tau_{1,\n}^{2k_1k_2-1}\tau_{1,\n+2\e_2}^{k_2^2}}{\tau_{0,\n-\e_2}^{k_1}  \tau_{0,\n+\e_2}^{k_2}   }.\label{trans2}
\end{align}
Using these results,
\begin{align}
F_{0,\n}&=\frac{\tau_{2,\n}\tau_{0,\n}-\tau_{1,\n-\e_2}^{k_1}\tau_{1,\n+\e_2}^{k_2}}{\tau_{1,\n-\e_1}^{l_1}\tau_{1,\n+\e_1}^{l_2}}=\frac{  \tau_{1,\n-\e_2}^{k_1}\tau_{1,\n+\e_2}^{k_2} }{ \tau_{1,\n-\e_1}^{l_1}\tau_{1,\n+\e_1}^{l_2}  }(U_{1,\n}-1),\label{trans3}\\
F_{1,\n}&=\frac{  \tau_{2,\n-\e_2}^{k_1}\tau_{2,\n+\e_2}^{k_2} }{ \tau_{2,\n-\e_1}^{l_1}\tau_{2,\n+\e_1}^{l_2}  }(U_{2,\n}-1)\notag \\
&=\frac{ \tau_{0,\n-\e_1}^{l_1}\tau_{0,\n+\e_1}^{l_2}  }{  \tau_{0,\n-\e_2}^{k_1}\tau_{0,\n+\e_2}^{k_2} }
\frac{\tau_{1,\n-2\e_2}^{k_1^2}\tau_{1,\n}^{2k_1k_2-1}\tau_{1,\n+2\e_2}^{k_2^2}}{ \tau_{1,\n-2\e_1}^{l_1^2}\tau_{1,\n}^{2l_1l_2-1}\tau_{1,\n+2\e_1}^{l_2^2}  }\frac{  U_{1,\n-\e_2}^{k_1}U_{1,\n+\e_2}^{k_2} }{ U_{1,\n-\e_1}^{l_1}U_{1,\n+\e_1}^{l_2}  }(U_{2,\n}-1)
.\label{trans4}
\end{align}
\end{subequations}
The inverse mapping ($\bT\, \rightarrow \bW$) is constructed as
\begin{subequations}
\begin{equation}
\tau_{1,\n}=\frac{1}{\tau_{3,\n}}\left( F_{1,\n} \tau_{2,\n-\e_1}^{l_1}\tau_{2,\n+\e_1}^{l_2} +  \tau_{2,\n-\e_2}^{k_1}\tau_{2,\n+\e_2}^{k_2} \right),
\end{equation}
and
\begin{align}
\tau_{0,\n}&=\frac{1}{\tau_{2,\n}}\left( F_{0,\n} \tau_{1,\n-\e_1}^{l_1}\tau_{1,\n+\e_1}^{l_2} +  \tau_{1,\n-\e_2}^{k_1}\tau_{1,\n+\e_2}^{k_2} \right),\\
U_{2,\n}&=\frac{\tau_{3,\n}\tau_{1,\n}}{\tau_{2,\n-\e_2}^{k_1}\tau_{2,\n+\e_2}^{k_2}},\  U_{1,\n}=\frac{\tau_{2,\n}\tau_{0,\n}}{\tau_{1,\n-\e_2}^{k_1}\tau_{1,\n+\e_2}^{k_2}}.
\end{align}
\end{subequations}


\begin{thebibliography}{99}


\bibitem{Toda}
M. Toda: 
Vibration of a Chain with Non-linear Interaction,
\textit{J. Phys. Soc. Jpn.}, \textbf{22}, 431-436 (1967).


\bibitem{Mikhairov}
A.V. Mikhailov: Integrability of a two-dimensional generalization of the Toda chain, \textit{JETP Lett.}, \textbf{30}, 414--418 (1979).


\bibitem{Fordy-Gibbons}
A.P. Fordy and J. Gibbons: Integrable Nonlinear Klein-Gordon Equation and Toda Lattices, \textit{Comm. Math. Phys.}, \textbf{77}, 21--30 (1980).


\bibitem{Hirota}
R. Hirota, 
Nonlinear partial difference equations. II. Discrete-time Toda equation, J. Phys.
Soc. Jpn.,\textbf{43}, 2074--2078, (1977).


\bibitem{HTI}
R. Hirota, S. Tsujimoto, and T. Imai, Difference scheme of soliton equations, RIMS
Kokyuroku, \textbf{822}, 144-152 (1993).

\bibitem{Tsujimoto}
Satoshi Tsujimoto: Studies on Discrete Nonlinear Integrable Systems,
\textit{Doctor Thesis, Waseda University} (1997).


\bibitem{SC}
B. Grammaticos and A. Ramani and V. Papageorgiou,
Do integrable mappings have the
Painlev\'e property?,
Phys. Rev. Lett.,
\textbf{67},
1825--1828 (1991).


\bibitem{AE}
M.\ P.\ Bellon, C.-M. Viallet,
Algebraic Entropy,
Commun. Math. Phys. \textbf{204}, 425--437 (1999).


\bibitem{KMMT2}
M. Kanki, J. Mada, T. Mase and T. Tokihiro, 
Irreducibility and co-primeness as an integrability criterion for discrete equations
J. Phys. A, \textbf{47}, 465204 (2014).


\bibitem{dToda}
M. Kanki, J. Mada and T. Tokihiro,
Integrability criterion in terms of coprime property for the discrete Toda equation,
J. Math. Phys.,
\textbf{56},
022706 
 (2015).

\bibitem{KKMT}
R. Kamiya, M. Kanki, T. Mase and T. Tokihiro,
Coprimeness-preserving non-integrable extension to the two-dimensional discrete Toda lattice equation,
\textit{J. Math. Phys.}, \textbf{58}, 012702 (2017).




\bibitem{SOMOS}
D. Gale, 
The Strange and Surprising Saga of the Somos Sequences,
Math. Intel., \textbf{13},  40--42 (1991).


\bibitem{KMT}
M. Kanki, T. Mase and T. Tokihiro,
Singularity confinement and chaos in two-dimensional discrete systems
J. Phys. A, \textbf{49}, 23LT01 (2015).



\bibitem{RGH}
 A. Ramani and B. Grammaticos and J. Hietarinta,
Discrete versions of the Painlev\'e equations,        
    Phys. Rev. Lett.,
    \textbf{67},
    1829--1832  (1991).

\bibitem{Sakai}
H. Sakai,
Rational surfaces associated with affine root systems and geometry of the Painlev\'{e} equations,
 Commun. Math. Phys. \textbf{220}, 165–229 (2001).



\bibitem{HV}
  J. Hietarinta and C. Viallet,
Singularity confinement and chaos in discrete systems,
    Phys. Rev. Lett.,
   \textbf{81},
   325--328 (1998).





\end{thebibliography}
\end{document}